\documentstyle[aps,epsfig,prl,multicol]{revtex}

\begin{document}
\title{
Heat conductivity in linear mixing systems
}
\author{
Baowen Li$^a$, Giulio Casati$^{a,b}$, and Jiao Wang$^a$ }
\address{
$^a$Department of Physics, National University of Singapore,
117542 Singapore \\
$^b$
International Center for the Study of Dynamical Systems, Universita' degli studi dell'Insubria, via Valleggio,11
22100 Como, and \\
Istituto Nazionale di Fisica della Materia, Unita' di Como, and \\
Istituto Nazionale di Fisica Nucleare, sezione di Milano, Via Celoria, 16
Milano, Italy
}
\date{\today}
\maketitle

\begin{abstract}

We present analytical and numerical results on the heat conduction in a linear
mixing system. In particular we consider a quasi one dimensional channel with
triangular scatterers with internal angles irrational multiples of $\pi$ and we
show that the system obeys Fourier law of heat conduction. Therefore deterministic
diffusion and normal heat transport which are usually associated to full hyperbolicity,
actually take place in systems without exponential instability.

\end{abstract}

\pacs{PACS numbers: 44.10.+i,  05.45.-a, 05.70.Ln, 66.70.+f}
\begin{multicols}{2}

Given a particular classical, many-body Hamiltonian system,
neither phenomenological nor fundamental transport theory can
predict whether or not this specific Hamiltonian system yields an
energy transport governed by the Fourier heat law\cite{Peie}. Heat
flow is universally presumed to obey a simple diffusion equation
which can be regarded as the continuum limit of a discrete random
walk. In consequence, transport theory requires that the
underlying deterministic dynamics yield a truly random process.
Therefore, it is not mere idle curiosity to wonder what class, if
any, of many-body systems satisfy the necessary stringent
requirements. Moreover, it now becomes increasingly meaningful to
seek example of many-body systems which, using dynamics alone, can
be shown to obey the Fourier heat law. A large number of papers
have recently approached this problem, mainly via numerical
simulations
\cite{Bon,Ford,Prosen,Kab93,LLP,HLZ,Hata,Gendelman,alonso,Aoki01}.
Leaving aside, for the purpose of the present paper, systems which
conserve the total momentum,  the general picture which emerges is
that positive Lyapounov exponent is a sufficient condition to
ensure Fourier heat law. In particular the paper \cite{alonso} was
precisely aimed at answering this question. Indeed in
\cite{alonso} the thermal conductivity was studied for a Lorentz
channel- a quasi one dimensional billiard with circular scatterers
- and it was shown to obey Fourier law. The conclusion at which
the above numerical computations point out appears quite natural.
Indeed modern ergodic theory tells us that for K-systems, a
sequence of measurements with finite precision mimics a truly
random sequence and therefore these systems appear precisely those
deterministically random systems tacitly required by transport
theory. On the other hand we do not have rigorous results and in
spite of several efforts, the connection between Lyapounov
exponents, correlations decay and diffusive properties is still
not completely clear. In a recent paper \cite{triangle} a model
has been presented which has zero Lyapounov exponent and yet it
exhibits unbounded Gaussian diffusive behavior. Since diffusive
behavior is at the root of normal heat transport then the above
results constitutes a strong suggestion that normal heat
conduction can take place even without the strong requirement of
exponential instability. If this would be the case, then it may
turn out to be an important step in the general attempt to derive
macroscopic statistical laws from the underlying deterministic
dynamics. Indeed systems with zero Lyapounov exponent have zero
algorithmic complexity and, at least in principle, are
analytically solvable.

\begin{figure}
\epsfxsize=8.cm \epsfbox{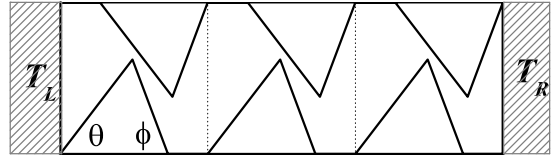}
 \vspace{0cm}
 \narrowtext
\caption {The geometry of the model. Particles move in the region
outside the triangular scatterers. The {\it x} coordinate goes
along the channel and {\it y } is perpendicular to it. The two
heat reservoirs at temperatures $T_L$ and $T_R$ are indicated. The
length of each cell is $l= 3$, the base of the triangles is $a=
2.19$ and the distance between the two parallel lines is $d=1.8$.
The geometry is then uniquely specified by assigning the internal
angles $\theta$ and $\phi$ } \label{model}
\end{figure}

In this paper we  consider  a two dimensional billiard model which
consists of two parallel lines of length $L$  at distance $d$  and
a series of triangular scatterers (Fig. 1). In this geometry, no particle can
move between the two reservoirs without suffering elastic
collisions with the triangles. Therefore this model is analogous
to the one studied in \cite{alonso} with triangles instead of
discs and the essential difference is that in the triangular model
discussed here the dynamical instability is linear and therefore
the Lyapounov exponent is zero. Strong numerical evidence has been
recently given \cite{casati} that the motion inside a triangular
billiard, with all angles irrational with $\pi$ is mixing, without
any time scale. Moreover, an area preserving map, which was
derived as an approximation of the boundary map for the irrational
triangle, when considered on the cylinder, shows a nice Gaussian
diffusive behavior even though the Lyapounov exponent of the map
is zero \cite{triangle}. It is therefore reasonable to expect that the motion
inside the irrational polygonal area of Fig 1 is diffusive thus
leading to normal conductivity.

In the following we present careful numerical investigations of
the heat transport in the  system of Fig 1 both by direct
numerical simulation of energy flow for the system in contact with
thermal baths as well as via Green-Kubo approach. Our results
provide convincing evidence that, if the angles $\theta$ and
$\phi$ are irrational multiples of $\pi$, the system
obeys Fourier law and the coefficients of thermal conductivity
computed via the two approaches, coincide.

In our computations we model the heat baths by stochastic kernels
of Gaussian type,  namely, the probability distribution of
velocities for particles coming out from the baths is

\begin{eqnarray}
P(v_x) =
\frac{|v_x|}{T}\exp\left(-\frac{v^2_x}{2T}\right),\nonumber\\
P(v_y) = \frac{1}{\sqrt{2\pi T}}\exp\left(-\frac{v^2_y}{2T}\right)
\label{Gaussian}
\end{eqnarray}
for $v_x$ and $v_y$, respectively.

\begin{figure}
\epsfxsize=8.cm \epsfbox{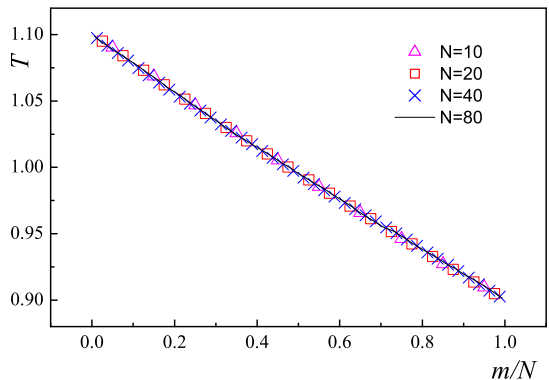}
\narrowtext \caption{Internal local temperature as a function of
the rescaled cell number $m/N$ for the irrational case with $\theta=(\sqrt{2}-1)\pi/2$ and $\phi=1$.
 The total number $N$ of cells is:
$N=10$ ($\triangle$), $N=20$ ($\Box$), $N=40$ ($\times$), and
$N=80$ (solid line). Here $T_L=1.1$, $T_R=0.9$, $l=3$. Notice the
quite good scaling behavior of the temperature field.}
\label{temperature}
\end{figure}

The total length of the channel is $L =Nl$ where $N$ and $l$ are
the number and the length of the fundamental cells. For the irrational
angles we take $\theta=(\sqrt{2}-1)\pi/2$ and $\phi=1$. By increasing
$L$, the number of particles per cell must be kept constant.
However, since the particles do not interact one may consider the
motion of a single particle over long times and then rescale the
flux.

 We turn now to the definition of the two relevant
quantities: the internal temperature and the heat flux. The
temperature field at the stationary state is calculated following
the idea used in Ref \cite{alonso}, namely, we divide the
configuration space in slices $\{C_i\}$. The time spent by the particle within the
slice in the $j$th visit is denoted by $t_j$ and the total number
of crossings of a slice $C_i$ during the simulation is $M$. The
temperature is defined by

\begin{equation}
T_{C_i}=\frac{\sum_j^Mt_jE_j(C_i)}{\sum_j^Mt_j},
\label{Temperature}
\end{equation}
where $E_j(C_i)$ is the kinetic energy at the $j$th crossing of
the slice $C_i$. Since the energy changes only at collisions with
the heat baths, we define the heat flux as

\begin{equation}
j (t_c)=\frac{1}{t_c}\sum_{k=1}^{N_c}(\Delta E)_k , \label{flux}
\end{equation}
where $(\Delta E)_k = E_{in} - E_{out}$ is the change of energy at
the $k$th collision with the heat bath and $N_c$ is the total number of
such collisions which occur during time $t_c$.

\begin{figure}
\epsfxsize=8.cm \epsfbox{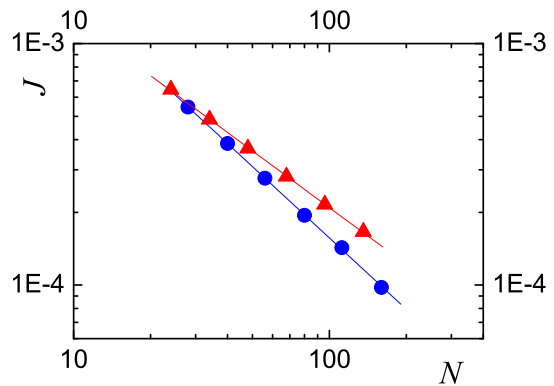} \narrowtext \caption{Scaling
behavior of the stationary heat flux $J$ as a function of the
system size for the irrational case of Fig. 1 ($\bullet$) and for
the rational case (solid triangle)(see later in the text). $N$ is
the number of fundamental cells. The best-squares fitting gives a
slope $ -0.99 \pm 0.01$ for the irrational case and $-0.78\pm0.01$
for the rational one.} \label{heatflux}
\end{figure}

For sufficiently long integration times, both the internal
temperature field and the heat flux reach a stationary value. We
have checked that the temperature profile obeys the law given in
Ref\cite{alonso}. For small temperature
differences $\Delta T$, it is a linear function, as illustrated in Fig. 2:

\begin{equation}
\nabla T =\frac{T_R- T_L}{L}.
 \label{Tgrad}
\end{equation}

In order to study the dependence of the heat flux on the system
size, we need to consider larger and larger systems, keeping the
particles density constant. The corresponding heat flux (for a
density of one particle per unit length) is $J = Lj$, where $j$ is
the flux computed with a single particle simulation. In Fig. 3, we
plot the heat flux $J$ as a function of the system size $N$. For
the irrational case, the best fit gives $J=AN^{-\gamma}$, with
$\gamma = 0.99\pm 0.01$ and $A=0.015$. The coefficient of thermal
conductivity is therefore independent on $N$, which means that the
Fourier law is obeyed and its numerical value is $\kappa
=-\frac{J}{\nabla T}= 0.225$.

The point under discussion here is very delicate and numerical
experiments must be very accurate and reliable in order to reach
clear conclusions. We have therefore checked the validity of our
result by an independent approach, via a Green-Kubo type
formalism, by studying the diffusive properties of our model
isolated from thermal baths.

\begin{figure}
\epsfxsize=8.cm \epsfbox{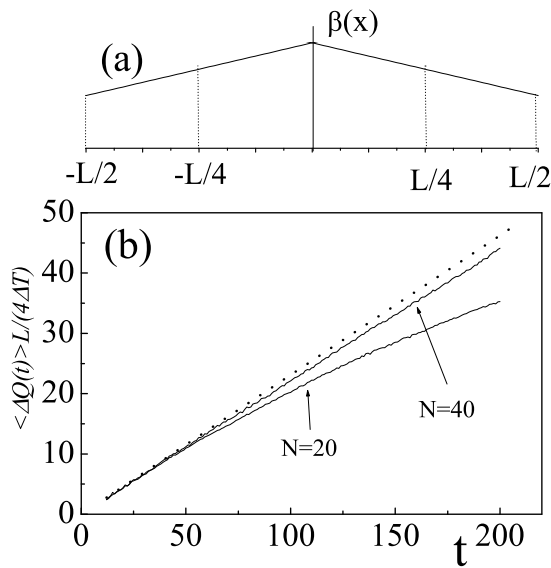}
 \narrowtext \caption {(a)
Initial temperature distribution $\beta(x) = 1/T(x)$ with
$\beta(0)=1.1, \beta(\pm L/2) =0.9$. (b)The rescaled heat flow
$\frac{\langle \Delta Q(t)\rangle }{2 T^2 \nabla\beta}$ versus
time $t$ for different values of chain length for the irrational
case. The dotted line has slope 0.225.} \label{conductivity}
\end{figure}

In our numerical calculations, we follow Ref\cite{Ford}, namely,
we consider the system with periodic boundary conditions and with
an initial temperature distribution $\beta(x)$ given in Fig. 4(a).
Then we calculate how the heat flows from the half hotter part
($L/4<|x|<L/2$) of the system to the half colder part ( $-L/4 <x<
+L/4$). At time $t=0$ we take a Maxwellian distribution of
velocities, namely, $P(v_{x,y})=
\exp(-v^2_{x,y}/(2T(x)))/\sqrt{2\pi T(x)}$. If we denote by $Q(t)$
the energy contained in the cold half part of the chain then,  if
the system obeys the Fourier law, the quantity ($\langle\Delta
Q(t)\rangle = \langle Q(t) -Q(0)\rangle $) must increase linearly
with time $t$, $\langle\Delta Q(t)\rangle= (2 \kappa T^2 \nabla
\beta) t$. Clearly the linear increase takes places only for times
smaller than the sound transit time across channel. The numerical
results are shown in Fig. 4(b) where the dotted line, which fits
the initial linear increase of the curve for N=40, has slope $
\kappa = 0.225$ thus indicating a very good agreement with
simulations with thermal baths.

Another important characteristic which is relevant for transport
properties  is the decay of the velocity autocorrelation function.
In Fig 5 we show the decay of the absolute value of the normalized
velocity autocorrelation $C(t) =\frac{\langle v_x(0)v_x(t)\rangle}
{\langle v^2_x(0)\rangle}$ for the irrational case. This function
decays as $|C(t)| \sim t^{-\alpha}$ with $\alpha \approx 3/2$.

\begin{figure}
\epsfxsize=8.cm \epsfbox{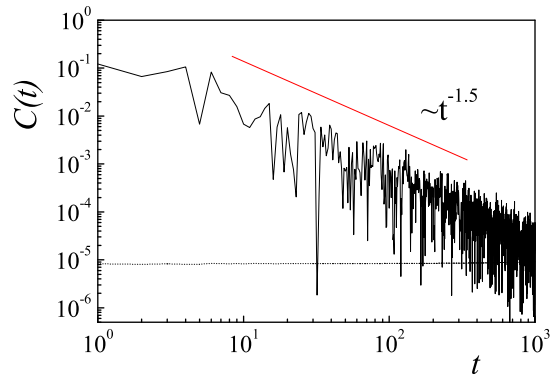} \vspace{0cm} \narrowtext
\caption {The absolute value of the velocity autocorrelation
function $|C(t)|$ averaged over $6\times 10^5$ orbits initially
with unit velocity amplitude but random directions. The solid line
has slope $-3/2$. The dotted line shows the statistical errors. }
\label{correlation}
\end{figure}

In Fig. 6 we show the diffusive behaviour of our model isolated
from thermal baths. A clear diffusive behaviour $\langle \Delta
x^2\rangle = 2Dt$ for the irrational case is observed with diffusion
coefficient $D = 0.15$.

In fact, in our model, we can establish a connection between the
thermal conductivity $\kappa$ and the diffusion coefficient $D$.
Indeed the average time a particle takes to travel from the left
to the right bath and vice versa is:
\begin{equation}
\langle t_{LR}\rangle = \langle t_{RL}\rangle=\frac{L^2}{2D},
\end{equation}
then according to Eq.(\ref{flux}), we can write the heat current
as:
\begin{equation}
j = \frac{\int_0^{\infty}
\frac{v^2}{2}\left(P(v,T_L)-P(v,T_R)\right)dv}{\langle
t_{LR}\rangle+\langle t_{RL}\rangle},
 \label{flux2}
\end{equation}
where  $P(v,T)= (4\pi/(2\pi T)^{3/2}) v^2 exp (-v^2/2T)$ is the
distribution for the modulus of velocity of particles coming out
from the baths (see Eq. (1)). The thermal conductivity $\kappa = -
Lj/\nabla T$ is thus,

\begin{equation}
\kappa=\frac{3}{2} D.
 \label{KappaD}
\end{equation}

We have numerically tested this formula. The thermal conductivity
calculated from Figs.2-3 is $\kappa=0.225$. On the other hand the
diffusion coefficient calculated from Fig. 6 is $D=0.15$ which,
according to the above relation, gives $\kappa = 3D/2 = 0.225$.

As expected, a completely different behaviour is obtained when the
angles $\theta$ and $\phi$ are rational multiples of $\pi$. The case with
$\theta=\pi/5$ and $\phi=\pi/3$, is shown in Fig. 6 (triangles),
and leads to a clear anomalous diffusive behaviour indicating the
absence of Fourier law. Correspondingly, the simulations with the
heat baths (solid triangles in Fig. 4) indicate a divergent
behaviour of the coefficient of thermal conductivity $ \kappa \sim
N^{0.22}$.

\begin{figure}
\epsfxsize=8.cm \epsfbox{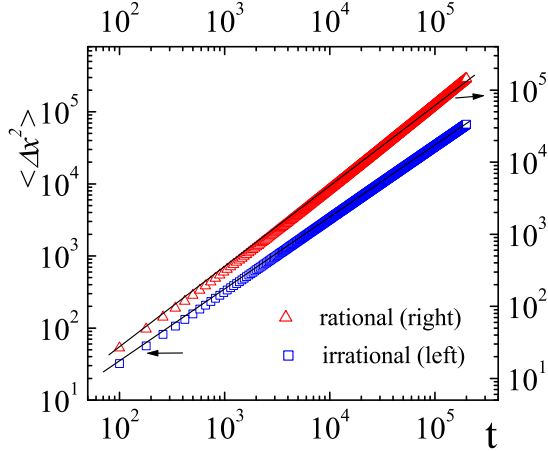} \vspace{0.cm} \narrowtext
\caption {Diffusive properties of our model isolated from thermal baths. In the irrational
case $\langle \Delta x^2\rangle =0.308 t^{1.007}$ ($\Box$); In the
rational case,  $\langle \Delta x^2\rangle =0.082
t^{1.178}$($\triangle$). $\Delta x^2\equiv (x(t)-x(0))^2$. In the
numerical calculation, $4\times 10^5$ particles are used. The
particles are initially at $x=0$ (in the center of the chain) and the initial velocities obey the
Maxwell-Boltzman distribution at temperature $T=1$.  }
\label{diffusion}
\end{figure}

In conclusion, when all angles are irrational multiples of $\pi$
the model shown in Fig. 1 exhibits Fourier law of heat conduction
together with nice diffusive properties and
the numerical value of the thermal conductivity computed via a
Green-Kubo approach agrees with the one obtained by direct
numerical simulations with thermal baths. However, when all angles
are rational multiple of $\pi$, the  model shows abnormal
diffusion and the heat conduction does not follow the Fourier law.

One may argue that the model considered here is somehow artificial
and far from realistic physical models. However the problem
discussed here is quite delicate and controversial and it is
highly desirable to understand which dynamical properties are
necessary and sufficient to derive Fourier law. In this respect
billiards like models are very convenient since they are more
suitable for analytical and numerical analysis.

\bigskip
BL and JW were supported in part by Academic Research Fund of NUS.
GC has been partially supported by EU Contract No.
HPRN-CT-1999-00163 (LOCNET network) and by MURST (Prin 2000,
\emph{Caos e localizzazione in Meccanica Classica e Quantistica}).

\end{multicols}

\end{document}